\documentclass[twoside]{article}
\usepackage{fleqn,espcrc2,axodraw,amsfonts,latexsym}

\usepackage{graphicx}
\usepackage[figuresright]{rotating}

\newcommand{\AmS}{{\protect\the\textfont2
  A\kern-.1667em\lower.5ex\hbox{M}\kern-.125emS}}

\title{U(1) Chiral gauge theory on lattice with gauge-fixed domain wall fermions\thanks{Combination of two talks by the authors}}

\author{S. Basak\address{Theory Group, Saha Institute of Nuclear Physics, 1/AF, Salt Lake, 		Calcutta 700064, India}%
	\thanks{Permanent address: N.N.D. College, Calcutta 700092, India.}
        and
        Asit K. De\addressmark}
       
\begin{document}

\begin{abstract}
We investigate a U(1) lattice chiral gauge theory ($ L\chi GT$) with domain wall fermions and gauge fixing. In the {\em reduced} model limit, our perturbative and numerical investigations at Yukawa coupling $y=1$ show that there are no extra mirror chiral modes. The longitudinal gauge degrees of freedom have no effect on the free domain wall fermion spectrum consisting of opposite chiral modes at the domain wall and the anti-domain wall which have an exponentially damped overlap. Our numerical investigation at small Yukawa couplings ($y\ll 1$) also leads to similar conclusions as above.
\vspace{1pc}
\end{abstract}

\maketitle

\section{Introduction}
Lattice regularization of chiral gauge theories has remained a long standing
problem of nonperturbative investigation of quantum field theory. Lack of
chiral gauge invariance in $L\chi GT$ proposals is responsible for
the longitudinal gauge degrees of freedom ({\em dof}) coupling to
fermionic {\em dof} and eventually spoiling the chiral nature of the theory.
The obvious remedy to control the longitudinal gauge {\em dof}
is to gauge fix with a target theory in mind \cite{golter1}. The formal
problem is that for compact
gauge-fixing a BRST-invariant partition function as well as (unnormalized)
expectation values of BRST invariant operators vanish as a consequence of
lattice Gribov copies \cite{neuberger}. Shamir and Golterman \cite{golter1}
have proposed to
keep the gauge-fixing part of the action BRST non-invariant and tune
counterterms to recover BRST in the continuum. In their formalism, the
continuum limit is to be taken from within the broken ferromagnetic (FM) phase
approaching another broken phase which is called ferromagnetic directional
(FMD) phase, with the mass of the gauge field vanishing at the FM-FMD
transition. This was tried out in a U(1) Smit-Swift model and so
far the results show that in the pure gauge sector, QED is recovered in
the continuum limit \cite{recent} and in the {\em reduced} model limit free
chiral fermions in the appropriate chiral representation are obtained
\cite{bock1}.

When one gauge transforms a gauge non-invariant theory, one picks up the longitudinal gauge degrees of freedom (radially frozen scalars) explicitly in the action. The reduced model is then obtained by making the lattice gauge field unity for all links, {\em i.e.}, by switching off the transverse gauge coupling. The reduced model is obviously a Yukawa model. The job of gauge fixing in the $L\chi GT$ when translated into the reduced model is to find a continuous phase transition where these unwanted scalars are decoupled leaving only free fermions in the appropriate chiral representation.  

We want to apply the gauge-fixing proposal to other previous proposals of a $L\chi GT$ which
supposedly failed due to lack of gauge invariance. For this purpose we have
chosen the waveguide formulation of the domain wall fermion where, without
gauge fixing, mirror chiral modes appeared at the waveguide boundary in
addition to the chiral modes at the domain wall or anti-domain wall to spoil
the chiral nature of the theory \cite{golter2}. Our investigation of the reduced model in this case reveals that both at Yukawa coupling $y=1$ \cite{prd} and $y\ll 1$ \cite{plb} the scalars are fully decoupled, there are no mirror modes and the spectrum is that of free domain wall fermions.

\section{Gauge-fixed domain wall action}
For Kaplan's free domain wall fermions \cite{kaplan} on a $4+1$-dimensional
lattice of size $L^4 L_s$ where $L_s$ is the 5th dimension, with periodic
boundary conditions in the 5th or $s$-direction and the domain wall mass
$m(s)$ taken as
\[ m(s) = \begin{array}{cl}
-m_0 & ~~~~~0 <s< \frac{L_s}{2} \\
 0   & ~~~~~s = 0, \frac{L_s}{2} \\
 m_0 & ~~~~~\frac{L_s}{2} <s< L_s
\end{array} \]
the model possesses a lefthanded (LH) chiral mode bound to the domain wall at
$s=0$ and a righthanded (RH) chiral mode bound to the anti-domain wall at
$s=\frac{L_s}{2}$. For $m_0 L_s\gg 1$, these modes have exponentially
small overlap.

A 4-dimensional gauge field which is same for all
$s$-slices can be coupled to fermions only for a restricted number of
$s$-slices around the anti-domain wall \cite{golter2} with a view to coupling
only to the RH mode at the anti-domain wall. The gauge field is thus confined
within a {\em waveguide}, $WG = (s: s_0 < s \leq s_1)$
with $s_0 = \frac{L_s+2}{4}-1$,
$s_1 = \frac{3L_s+2}{4}-1$. For convenience,
the boundaries at ($s_0,s_0+1$) and ($s_1,s_1+1$) are denoted waveguide
boundary $I$ and $II$ respectively.
The symmetries of the model remain exactly the same as in \cite{golter2}.

Obviously, the hopping terms from $s_0$ to $s_0+1$ and that from $s_1$ to
$s_1+1$ would break the local gauge invariance of the action. This
is taken care of by gauge transforming the action and thereby picking up the
pure gauge {\em dof} or a radially frozen scalar field $\phi$
(St\"{u}ckelberg field) at the $WG$ boundary, leading to the gauge-invariant
action (with $\varphi_x\rightarrow g_x \varphi_x$, where $g\in$ gauge group):
\begin{eqnarray}
S_{F} & = & \sum_{s \in WG}
\overline{\psi}^s \left( D\!\!\!\!/~(U) - W(U) + m(s) \right) \psi^s
\nonumber \\
&+& \sum_{s\not\in WG} \overline{\psi}^s \left( \partial\!\!\!/
- w + m(s) \right) \psi^s + \sum_s \overline{\psi}^s \psi^s \nonumber \\
&-& \sum_{s\neq s_0,s_1}
\left( \overline{\psi}^s P_L \psi^{s+1} + \overline{\psi}^{s+1} P_R \psi^s
\right) \nonumber \\
&-& y \left( \overline{\psi}^{s_0} \phi^\dagger P_L \psi^{s_0+1} +
\overline{\psi}^{s_0+1} \phi P_R \psi^{s_0} \right) \nonumber \\
&-& y \left( \overline{\psi}^{s_1} \phi P_L
\psi^{s_1+1} + \overline{\psi}^{s_1+1} \phi^\dagger P_R \psi^{s_1} \right)
\label{wgact}
\end{eqnarray}
where we have taken the lattice constant $a=1$
and have suppressed all other indices than $s$. The projector $P_{L(R)}$
is $(1\mp \gamma_5)/2$ and $y$ is the Yukawa coupling at the $WG$ boundaries.
The $D\!\!\!\!/~(U)$ and $W(U)$ are the gauge covariant Dirac operator and
the Wilson term (with Wilson $r=1$) respectively. $\partial\!\!\!/$ and $w$ are the free versions of
$D\!\!\!\!/~(U)$ and $W(U)$ respectively.

The gauge-fixed pure gauge action for U(1), where the
ghosts are free and decoupled, is:
\begin{equation}
S_B(U) = S_g(U) + S_{gf}(U) + S_{ct}(U) \label{ggact}
\end{equation}
where, $S_g$ is the usual Wilson plaquette action;
the gauge fixing term $S_{gf}$ and the gauge field mass
counter term $S_{ct}$ are given by (for a discussion of relevant counterterms
see \cite{golter1,bock2}),
\begin{eqnarray}
S_{gf}(U) & = & \tilde{\kappa} \left( \sum_{xyz} \Box(U)_{xy}
\Box (U)_{yz} - \sum_x B_x^2 \right) \label{gfact} \\
S_{ct}(U)  & = &  - \kappa \sum_{x\mu} \left( U_{\mu x} +
U_{\mu x}^\dagger \right),
\end{eqnarray}
where $\Box(U)$ is the covariant lattice laplacian and
\begin{equation}
B_x  = \sum_\mu \left( \frac{V_{\mu x-\hat{\mu}} +
V_{\mu x}}{2} \right)^2  \label{bxsq}
\end{equation}
with
$V_{\mu x} = \frac{1}{2i} \left( U_{\mu x} - U_{\mu x}^\dagger \right)$
and $\tilde{\kappa} = 1/(2\xi g^2)$.
$S_{gf}$ has a unique absolute minimum at $U_{\mu x}=1$, validating
weak coupling perturbation theory (WCPT) around $g=0$ or
$\tilde{\kappa}=\infty$ and in the naive continuum limit it reduces to
$\frac{1}{2\xi} \int d^4x (\partial_\mu A_\mu)^2$.

Obviously, the action $S_B(U)$ is not gauge invariant. By giving it a gauge
transformation the resulting action $S_B(\phi^\dagger_x U_{\mu x}
\phi_{x+\hat{\mu}})$ is gauge-invariant with $U_{\mu x}\rightarrow g_x U_{\mu
x} g^\dagger_{x+\hat{\mu}}$ and $\phi_x \rightarrow g_x \phi_x$, $g_x \in
U(1)$. By restricting to the trivial orbit, we arrive at the so-called
{\bf reduced model} action
\begin{equation}
S_{reduced} = S_F(U=1) + S_B(\phi^\dagger_x \;1 \;\;\phi_{x+\hat{\mu}})
\label{reduced}
\end{equation}
where $S_F(U=1)$ is obtained quite easily from eq.(\ref{wgact}) and
\begin{eqnarray}
\lefteqn{S_B(\varphi^\dagger_x \;1 \;\;\varphi_{x+\hat{\mu}})
= -\kappa \sum_x \varphi^\dagger_x(\Box \varphi)_x}\nonumber \\ & & + \tilde{\kappa}
\sum_x \left[\varphi^\dagger_x(\Box^2 \varphi)_x - B^2_x \right] \label{redB}
\end{eqnarray}
now is a higher-derivative scalar field theory action. $B_x$ in (\ref{redB})
is same as in (\ref{bxsq}) with
\begin{equation}
V_{\mu x} = \frac{1}{2i}\left(\varphi^\dagger_x\varphi_{x+\hat{\mu}} -
\varphi^\dagger_{x+\hat{\mu}} \varphi_x\right).
\end{equation}

%%%%%%%%%%%%
\section{Weak Coupling Perturbation Theory in the reduced model}
%%%%%%%%%%%%%%%%%%%%%%%%%%%%%%%%%%%%%%%%%%%%%%%%%%%%%%%%%%%%%%%%
At $y=1$, we carry out a WCPT in the coupling $1/\tilde{\kappa}$ for the
fermion propagators to 1-loop. 
\subsection{Free propagators}
In order to develop perturbation theory,
in reduced model,we expand, $\phi_x = \exp(ib\theta_x)$ where
$b=1/\sqrt{2\tilde{\kappa}}$, leading to free propagator for the compact scalar $\theta$
\cite{bock2},
\begin{equation}
{\cal G}(p) = \frac{1}{\hat{k^2}(\hat{k^2} + \omega^2)},
~~~~~~~~\omega^2=\frac{\kappa}{\tilde{\kappa}}
\end{equation}
where, $\hat{k}_\mu = 2\sin(k_\mu/2)$. 

Free fermion propagators at $y=1$
are obtained in momentum space for 4-spacetime dimensions while staying
in the coordinate space for the 5th dimension following \cite{aoki}
(results in \cite{aoki} cannot be directly used because of difference in
finer details of the action). The appropriate free action to start from is,
\begin{equation}
S^0 = \sum_{p,s,t} \overline{\tilde{\psi}}_p^s \left[
i\overline{p\!\!\!/} \delta_{s,t} + MP_L + M^\dagger P_R
\right] \tilde{\psi_p^t}
\end{equation}
where, $M = F(p)\delta_{s,t} + M_0$, $M_0 = a(s)\delta_{s,t}
- \delta_{s+1,t}$,
$a(s) = 1 + m(s)$, $F(p) = \sum_\mu (1 - \cos(p_\mu))$,
$\overline{p}_\mu = \sin(p_\mu)$ and $\overline{p\!\!\!/} =
\gamma_\mu\overline{p}_\mu$.
The free fermion propagator can formally be written as,
\begin{eqnarray}
\lefteqn{\Delta(p)=\left[ i\overline{p\!\!\!/} + M P_L +
                                  M^\dagger P_R \right]^{-1}} \nonumber \\
 & & = ( -i\overline{p\!\!\!/} + M^\dagger ) P_L G^L+
   ( -i\overline{p\!\!\!/} + M ) P_R G^R  \label{fprop}
\end{eqnarray}
Explicit solution of $G^{L(R)}(p)$ are obtained from
\begin{eqnarray}
\lefteqn{\left[ \overline{p}^2+1+B(s)^2 \right] G^L_{s,t} - B(s+1) G^L_{s+1,t}} \nonumber \\
&  &- B(s) G^L_{s-1,t} =  \delta_{s,t} \label{green}
\end{eqnarray}
and a similar equation for $G^R$. In (\ref{green}), $B(s)= F(p)+1+m(s)$.
We show only the calculations for obtaining $G^L$ and henceforth drop the
superscript $L$. Setting $G = G^-$ in the region $0 <s\leq L_s/2-1$ where
$B(s)=F(p)+1-m_0=a_{-}$ and $G = G^+$ in the
region $L_s/2 <s\leq L_s-1$ where $B(s)=F(p)+1+m_0=a_{+}$,
\begin{eqnarray}
\lefteqn{\left( \overline{p}^2+1+a^2_{-} \right) G^{-}_{s,t} - a_{-} G^{-}_{s+1,t}
- a_{-} G^{-}_{s-1,t}} \nonumber \\  & & = \delta_{s,t} \label{lmslice} \\
\lefteqn{\left( \overline{p}^2+1+a^2_{+} \right) G^{+}_{s,t} - a_{+} G^{+}_{s+1,t}
- a_{+} G^{+}_{s-1,t}} \nonumber \\ & & = \delta_{s,t}. \label{lpslice}
\end{eqnarray}
The solutions of these equations are expressed as sum of homogeneous
and inhomogeneous solutions:
\begin{eqnarray}
\lefteqn{G^{\pm}_{s,t}(p)  =  g_{\pm}^{(1)}(t)e^{-\alpha_{\pm}(p)s} +
                     g_{\pm}^{(2)}(t)e^{\alpha_{\pm}(p)s}} \nonumber \\
 & & + \frac{\cosh[\alpha_{\pm}(p)(\vert s-t \vert - l/2)]}
{2a_{\pm}\sinh(\alpha_{\pm}(p))\sinh(\alpha_{\pm}(p)l/2)}. \label{lmsol}
\end{eqnarray}
where, $\cosh(\alpha_{\pm}(p))=\frac{1}{2}\left(
a_{\pm}+\frac{1+\overline{p}^2} {a_{\pm}}
\right)$.
The third term in the above is the inhomogeneous solution.
In order to get the complete solution we need to determine the
unknown functions $g_{\pm}^{(1)}(t)$ and $g_{\pm}^{(2)}(t)$ in (\ref{lmsol}),
which are obtained by considering boundary conditions from eqs.
(\ref{lmslice}) at $s=0,~1,~L_s/2-1,~L_s/2,~L_s/2+1,~L_s-1$,

\begin{eqnarray}
&  a_0 G^{-}_{0,t}(p) &= a_{+} G^{+}_{L_s,t}(p), \nonumber \\
&  a_0 G^{+}_{L_s/2,t}(p) &= a_{-} G^{-}_{L_s/2,t}(p), \nonumber \\
&  (\overline{p}^2+1+a_0^2)&G^{-}_{0,t}(p) - a_{-}G^{-}_{1,t}(p) \nonumber \\
&  &=\delta_{0,t} + a_0 G^{+}_{L_s-1,t}(p), \nonumber \\
&  (\overline{p}^2+1+a_0^2)&G^{+}_{L_s/2,t}(p) - a_{+}G^{+}_{L_s/2+1,t}
\nonumber \\
&  &= \delta_{L_s/2,t} + a_0 G^{-}_{L_s/2-1,t}. \label{dwbc}
\end{eqnarray}
with $B(s)=F(p)+1=a_0$ at $s=0,L_s/2$. Using the boundary conditions
(\ref{dwbc}) we arrive at an equation of the form, \begin{equation}
{\bf A}\cdot{\bf g}(t) = {\bf X}(t), \label{mateq}
\end{equation}
where, ${\bf g}(t) = (g_{-}^{(1)}~~~g_{-}^{(2)}~~~g_{+}^{(1)}~~~g_{+}^{(2)})$
is a 4-component vector, ${\bf X}(t)$ is another 4-component vector and ${\bf
A}$ is a $4\times 4$ matrix.

The solution to the equations (\ref{mateq}) is very complicated in general,
particularly for finite $L_s$, however $g_{\pm}(t)$ can be obtained by solving
the above equations numerically for different $t$ values. This way we can
easily construct the free chiral propagators at any given
$s$-slice, including $s=t=0,L_s/2$. The solutions for $G^R_{s,t}(p)$ and
the resulting propagators can be obtained in a similar way.

\subsection{Tree level fermion mass matrix}
Another issue of interest at the free fermion level is the spread of the
wavefunctions of the two chiral zero mode solutions along the discrete
$s$-direction and their possible overlap. A finite overlap would
mean an induced Dirac mass. The extra dimension can be interpreted as a flavor
space with one LH chiral fermion, one RH chiral fermion and $(L_s/2-1)$
heavy fermions on each sector of $1\leq s\leq L_s/2-1$ and $ L_s/2+1\leq
s\leq L_s-1$.  We consider
flavor diagonalization of $M^\dagger_0 M_0$ ($M_0=M(p=0)$ is not hermitian):
\begin{eqnarray}
\left( M^\dagger_0 M_0 \right)_{st} \left( \phi^{(0)}_j \right)_t &=& \left[
\lambda_j^{(0)} \right]^2 \left( \phi^{(0)}_j \right)_s \\
\left( M_0 M^\dagger_0 \right)_{st} \left( \Phi^{(0)}_j \right)_t &=& \left[
\lambda_j^{(0)} \right]^2 \left( \Phi^{(0)}_j \right)_s, \label{eveq}
\end{eqnarray}
where the index $j$ for the eigenvalues and the eigenvectors is a
flavor index.

We have carried out explicit solutions for the chiral modes at the domain and anti-domain walls and for the heavy modes \cite{prd}. We do not present these results explicitly here except to point out that the overlap of the opposite chiral modes at the domain and anti-domain walls is exponentially damped anywhere in the $s$ direction.

\subsection{1-loop fermion self-energy}

Next we calculate the fermion propagators to 1-loop. A {\em half-circle}
diagram which is diagonal in flavor space contributes to $LL$ and $RR$
propagator self-energies and a flavor off-diagonal {\em tadpole} diagram
produces the self-energy for the $LR$ and $RL$ parts. However, the
self-energies are nonzero only at the waveguide boundaries $I$ and $II$.
Besides, there is also a contribution to the fermion self-energy for the 
$LR$ and $RL$ parts coming from a flavor off-diagonal half-circle 
diagram (we shall call it a {\em global-loop} diagram) where the scalar 
field goes around the flavor space connecting fermions at the waveguide 
boundaries $I$ and $II$.  

By expanding $\phi_x = \exp(ib\theta_x)$ and
retaining up to ${\cal O}(b^2)$ in the interaction term
we find the vertices necessary to calculate the fermion self-energy to
1-loop.

\vspace{-0.3cm}
\hspace*{-2.0cm}\begin{picture}(290,85)(-10,65)
\ArrowLine(75,85)(125,85)
\ArrowLine(125,85)(175,85)
\ArrowLine(175,85)(225,85)
\DashArrowArcn(150,85)(35,180,0){3}
\Text(80,92)[]{\footnotesize{$L$}}
\Text(92,76)[]{$p$, $s_0+1$}
\Text(125,92)[]{\footnotesize{$R$}}
\Text(175,92)[]{\footnotesize{$R$}}
\Text(150,76)[]{$p-k$, $s_0$}
\Text(220,92)[]{\footnotesize{$L$}}
\Text(208,76)[]{$p$, $s_0+1$}
\Text(150,110)[]{$k$}
\end{picture}

The $LL$ propagator on the $(s_0+1)$-th slice at the waveguide boundary
$I$ receives a nonzero  self-energy contribution from the
half-circle diagram,
\begin{eqnarray*}
-\left(\Sigma_{LL}^I(p)\right)_{st} = \int_k
b^2 \left[-i\gamma_\mu (\overline{q})_\mu P_L
G_L(q)\right]_{s_0,s_0}
\end{eqnarray*}
\begin{equation}
~~~~~~~~~~~~~~~~~~~~~~~~~~
{\cal G}(k) \;\delta_{s,s_0+1}\delta_{t,s_0+1}
\label{fselfl} 
\end{equation}

\begin{equation} 
\rightarrow \, \frac{b^2}{L^4} \sum_k \left[{\cal
S}^{(0)}_{RR}(q)\right]_{s_0,s_0}{\cal G}(k) \;
\delta_{s,s_0+1}\delta_{t,s_0+1} \label{fselfd}
\end{equation}
where  $q=p-k$ and the expression in the square bracket in
(\ref{fselfl}) is the free $RR$ propagator 
$[{\cal S}^{(0)}_{RR}]_{s_0,s_0}$ on the $s_0$-slice. 
Eq.(\ref{fselfl}) assumes 
infinite 4 space-time volume while in eq.(\ref{fselfd}) a finite 
space-time volume $L^4$ is considered.  To avoid the infra-red 
problem in the scalar propagator, we use anti-periodic boundary 
condition in one of the space-time directions in evaluating 
(\ref{fselfd}). 

In a similar
way, 1-loop corrected $RR$ or $LL$ propagators are obtained at all the
$s$-slices of the waveguide boundaries $I$ and $II$, {\em i.e.}, at the
slices $s_0$, $s_0+1$, $s_1$ and $s_1+1$.

\vspace{-0.3cm}
\hspace*{-2.0cm}\begin{picture}(300,85)(0,65)
\ArrowLine(90,85)(150,85)
\ArrowLine(150,85)(210,85)
\DashArrowArcn(150,105)(20,180,0){3}
\DashCArc(150,105)(20,180,360){3}
\Text(95,92)[]{\footnotesize{$L$}}
\Text(112,76)[]{$p$, $s_0+1$}
\Text(205,92)[]{\footnotesize{$R$}}
\Text(198,76)[]{$p$, $s_0$}
\Text(150,115)[]{$k$}
\end{picture}

For the $LR$ propagator connecting $s_0$ and $s_0+1$ at the waveguide
boundary $I$, the self-energy contribution from the tadpole diagram is
given by,
\begin{eqnarray}
\lefteqn{-\left(\Sigma_{LR}^I(p)\right)_{st}}\nonumber \\ 
& & = \frac{1}{2}b^2P_L \int_k
\frac{1}{\hat{k}^2 (\hat{k}^2+\omega^2)}\;
\delta_{s,s_0}\delta_{t,s_0+1} \\
& & =\frac{1}{2}b^2P_L \,{\cal T}
\;\delta_{s,s_0}\delta_{t,s_0+1}  \label{tadpld}
\end{eqnarray}
where ${\cal T}\sim 0.04$ is the tadpole loop integral.

Similarly the self-energy contribution to the $LR$ propagator at the
waveguide boundary $II$ connecting $s_1$ and $s_1+1$ comes from a
tadpole diagram and is given by,
\begin{equation}
-\left(\Sigma_{LR}^{II}(p)\right)_{st}
= \frac{1}{2}b^2P_L \,{\cal T}
\;\delta_{s,s_1}\delta_{t,s_1+1}.  \label{tadpld2}
\end{equation}

The global-loop diagram originates from the fact that the $\varphi$ 
field that couples the fermions at the waveguide boundary $I$ is the 
same $\varphi$ field coupling the fermions at the waveguide boundary 
$II$.

\vspace{-0.3cm}
\hspace*{-2.0cm}\begin{picture}(300,85)(0,65)
\ArrowLine(62,85)(125,85)
\ArrowLine(125,85)(195,85)
\ArrowLine(195,85)(258,85)
\DashArrowArcn(160,85)(47,180,0){3}
\Text(67,92)[]{\footnotesize{$L$}}
\Text(82,76)[]{$p$, $s_0+1$}
\Text(125,92)[]{\footnotesize{$R$}}
\Text(125,76)[]{$s_0$}
\Text(195,92)[]{\footnotesize{$L$}}
\Text(156,76)[]{$p-k$}
\Text(193,76)[]{$s_1+1$}
\Text(253,92)[]{\footnotesize{$R$}}
\Text(247,76)[]{$p$, $s_1$}
\Text(160,121)[]{$k$}
\end{picture}

Self-energy contribution from the global-loop diagram for the $LR$ 
propagator is:  
\begin{eqnarray*}
-\left(\Sigma^{gl}_{LR}(p)\right)_{st} = b^2 P_L \int_k
\left[ M^\dagger(q) G_L(q) \right]_{s_1+1,s_0}
\end{eqnarray*}
\begin{equation}
~~~~~~~~~~~~~~~~~~~~~~~~~~~~~~~~~
{\cal G}(k) \;\delta_{s,s_1}\delta_{t,s_0+1} \label{global} 
\end{equation}
\begin{equation}
~~~~~~~~~~~~~~~~~~~
 = b^2 P_L \,{\cal R} \;\delta_{s,s_1}\delta_{t,s_0+1}  \label{nself}
\end{equation}
where $q=p-k$ and ${\cal R}$ is the loop integral in 
eq.(\ref{global}). 

The mass parameter $M_0$ gets modified to $\widetilde{M}_0$ at 1-loop as:
\begin{eqnarray}
(\widetilde{M}_0)_{st}P_L &=&
(M_0)_{st}P_L+\left[-(\Sigma^I_{LR}(0))_{st} \right] \nonumber \\ &+&
\left[-(\Sigma^{II}_{LR}(0))_{st}\right] +
\left[-(\Sigma^{gl}_{LR}(0))_{st}\right]  \nonumber \\
&\equiv & (M_0)_{st}P_L + b^2 ({\bf \Sigma}_{LR})_{st} P_L\,.
\end{eqnarray}
$\Sigma^{I,\,II}_{LL,RR}(0) =0$ identically. $(M_0^\dagger)_{st}P_R$
gets modified accordingly. 

\subsection{Fermion mass matrix diagonalization in 1-loop}
Diagonalization of the fermion mass matrix at 1-loop \cite{prd} shows that, i) the zero modes are perturbatively stable, and ii) the overlap of the opposite chiral modes at the domain and the anti-domain walls are still exponentially damped. This clearly rules out the necessity of a fermion mass counter term. 

%%%%%%%%%%%%%%%%%%%%%%%%%%%%%%%%%%%%%%%%%%%%%%
\section{Numerical results at $y=1$}
%%%%%%%%%%%%%%%%%%%%%%%%%%%%%%%%%%%%%%%%%%%%%%
\begin{figure}[t]
\vspace{0.2cm}
\includegraphics[width=6.5cm,height=5.0cm]{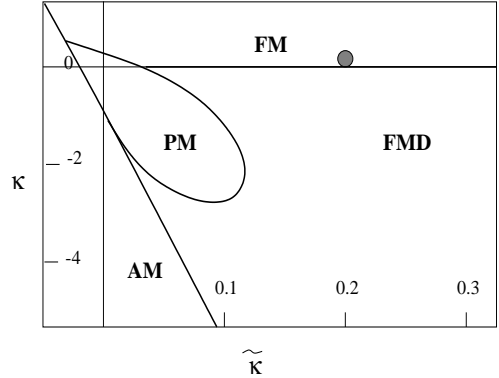}
\vspace{-0.5cm}
\caption{Schematic quenched phase diagram.}
\label{pd}
\vspace{-0.8cm}
\end{figure}

In the quenched approximation, we have first numerically confirmed the
phase diagram in \cite{bock3} of the reduced model in
($\kappa,\tilde{\kappa}$) plane. The phase diagram shown schematically in
Fig.1 has the interesting feature that for large enough $\tilde{\kappa}$,
there is a continuous phase transition between the broken phases FM and FMD.
FMD phase is
characterized by loss of rotational invariance and the continuum limit is to
be taken from the FM side of the transition. In the full theory with 
gauge fields, the gauge symmetry reappears at this transition and the gauge
boson mass vanishes, but the longitudinal gauge {\em dof} remain decoupled.
In Fig.1 PM is the symmetric phase and AM is the broken anti-ferromagnetic
phase. The numerical details involved in reconstruction of the phase 
diagram and the fermionic measurements that follow will be 
available in \cite{plb}. 

For calculating the fermion propagators, as in \cite{bock1} we have 
chosen the point $\kappa =0.05$, $\tilde{\kappa}=0.2$ (gray blob in 
Fig.1). Although this point is far away from $\tilde{\kappa}=\infty$, 
around which we did our perturbation theory in the previous section, the 
important issue here is to choose a point near the FM-FMD transition and 
away from the FM-PM transition. The results below show that for the 
fermion propagators there is 
excellent agreement between numerical results obtained at $\kappa =0.05$, 
$\tilde{\kappa}=0.2$ and perturbation theory. 
   
Numerically  on $4^3\times 16$ and $6^3\times 16$ 
lattices with $L_s=22$ and $m_0=0.5$ we look for chiral modes at the 
domain wall ($s=0$), the anti-domain wall ($s=11$), and at the waveguide 
boundaries ($s=5,6$ and $s=16,17$). Error bars in all the figures are 
smaller than the symbols. 

\begin{figure}[t]
\begin{center}
\vspace{-1.5cm}
\hspace*{-0.4cm}\includegraphics[width=8.5cm,height=10.5cm]{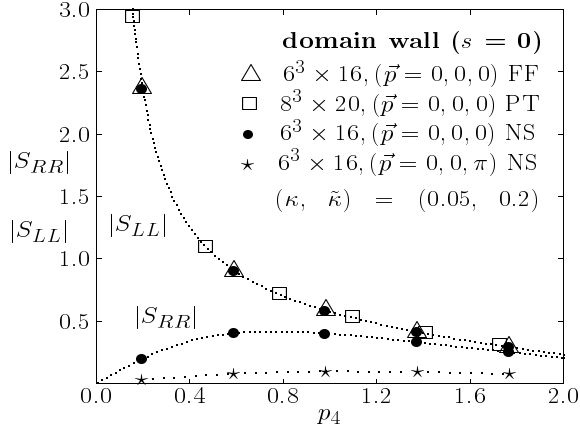}
\vspace{-4.2cm}
\caption{Propagators at domain wall $s=0$ ($L_s=22$; a.p.b.c. in $L_4$,
$y=1.0$).}
\label{lr0}
\end{center}
\vspace{-0.9cm}
\end{figure}

Figs.2 and 3 show the $RR$ propagator $|S_{RR}|$ and the $LL$ propagator
$|S_{LL}|$ at the domain and anti-domain wall as a function of a component of
momentum $p_4$ for both $\vec{p}=(0,0,0)$ (physical mode) and $(0,0,\pi)$
(first doubler mode) at $y=1$. From the figures, it is clear that
the doubler does not exist, only the physical $RR$ ($LL$) propagator seems to
have a pole at $p=(0,0,0,0)$ at the anti-domain (domain) wall. In all the
figures, NS, PT and FF respectively indicate data from numerical simulation,
from perturbation theory and from free fermion propagator by direct inversion
of the free fermion matrix.

For Figs.2 and 3, PT also mean zeroth order perturbation theory, {\em i.e.},
numerical solution of propagator following eq.(\ref{mateq}). We have PT
results also for $6^3\times 16$ lattice but have chosen not to show them
because they fall right on top of the numerical data. Instead PT results
shown for $8^3\times 20$ lattice for which the $p_4$ points are distinct. The
dotted line in all figures refer to the propagator from PT using a
$256^3\times 1024$ lattice. {\em The curves stay the same irrespective of
methods or lattice size}. Based on the above, we can conclude that there are
{\em only free RH fermions} at the anti-domain wall, and at the domain wall
there are {\em only free LH fermions}.

\begin{figure}[t]
\begin{center}
\vspace{-1.6cm}
\hspace*{-0.4cm}\includegraphics[width=8.5cm,height=10.5cm]{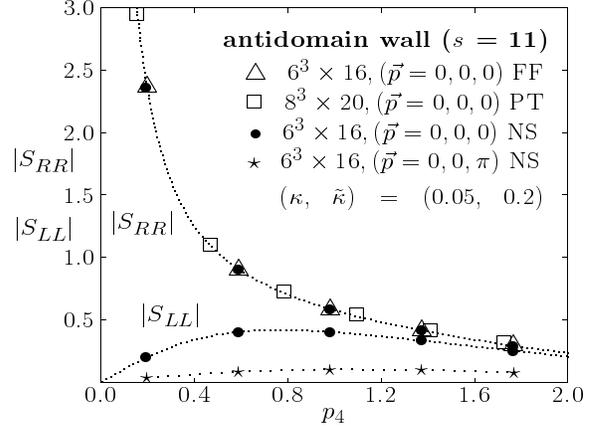}
\vspace{-4.0cm}
\caption{Chiral propagators at domain wall $s=0$ ($L_s=22$; a.p.b.c. in
$L_4$, $y=1.0$).}
\label{lr11}
\end{center}
\vspace{-0.9cm}
\end{figure}

Figs.4 and 5 show no evidence of a chiral mode at the waveguide boundaries
$s=5$ and 6 and excellent agreement with 1-loop perturbation
theory. Here too doublers do not exist. Actually the agreement with the FF
method (direct inversion of the free domain wall fermion matrix on a given
finite lattice) is also excellent, because the 1-loop corrections are almost
insignificant. For clarity, in Figs.4 and 5 we have not shown the
$LL$ propagator on $s=5$ and the $RR$ propagator at $s=6$, but conclusions are
the same.

\begin{figure}[t]
\begin{center}
\vspace{-1.6cm}
\hspace*{-0.4cm}\includegraphics[width=8.5cm,height=10.5cm]{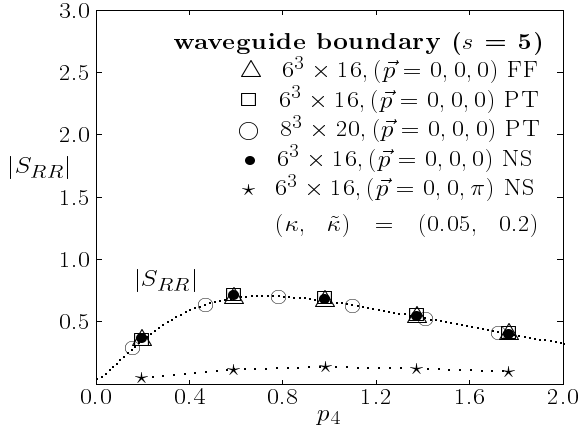}
\vspace{-4.0cm}
\caption{$RR$-propagator at waveguide boundary $s=5$ ($L_s=22$, a.p.b.c.
in $L_4$, $y=1.0$).}
\label{lr5}
\end{center}
\vspace{-0.9cm}
\end{figure}

\begin{figure}[t]
\begin{center}
\vspace{-1.6cm}
\hspace*{-0.4cm}\includegraphics[width=8.5cm,height=10.5cm]{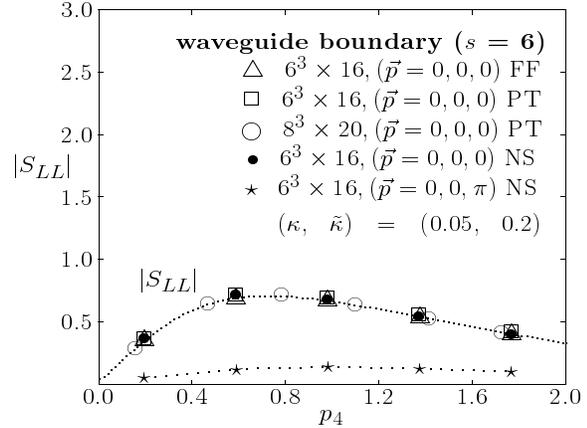}
\vspace{-4.0cm}
\caption{$LL$-propagator at waveguide boundary $s=6$ ($L_s=22$, a.p.b.c.
in $L_4$, $y=1.0$).}
\label{lr6}
\end{center}
\vspace{-0.9cm}
\end{figure}

Similar investigation at the other waveguide boundary $s=16, 17$ also does
not show any  chiral modes. Previous investigations of the domain wall 
waveguide model without gauge fixing \cite{golter2} have shown that the 
waveguide boundaries are the most likely places to have the unwanted 
mirror modes. This is why we have mostly concentrated in showing that 
there are no mirror chiral modes at these boundaries, although we have 
looked for chiral modes everywhere along the flavor dimension. In fact, 
we do not see any evidence of a chiral mode anywhere other than at the 
domain wall and the anti-domain wall.

\section{Numerical results at $y\ll 1$}
At $y=0$, the domain wall and the anti-domain wall are detached from each other. Fermion current considerations and numerical simulations clearly show in this case that mirror chiral modes form at the waveguide boundaries. 

On the other hand, at $y=1$ (as presented above) and $y>1$ \cite{bd99}, the mirror chiral modes are certainly absent at the waveguide boundaries. Actually in this case, the only chiral modes are at the domain wall and at the anti-domain wall and the spectrum is that of a free domain wall fermion.

The interesting question is what happens at $y$ smaller than unity, especially at positive values near zero. To investigate this question, we looked for chiral modes at $y=0.75,\,0.5,\,0.25$ at places other than the domain wall and the anti-domain walls, especially at the waveguide boundaries. On a $6^ 3\times 16$ lattice, the chiral propagators at the waveguide boundaries showed an increasing trend as the fourth component of momentum $p_4$ (with $\vec{p}=\vec{0}$) was decreased to the minimum value possible for this lattice size, something that could signify a pole at zero momentum. To resolve this we took lattice sizes which were bigger in the 4-direction to accommodate lower $p_4$-values. We found that for each $y<1$ there is a big enough lattice size for which the chiral propagators ultimately start showing a descending trend. Obviously the smaller the Yukawa coupling, the bigger the $L_4$ extension was required. Moreover, all the chiral propagators at these small Yukawa couplings matched exactly with the corresponding free case. Our results at the waveguide boundaries $s=5,\,6$ are summarized in the Figs.6 and 7 for the $RR$ and the $LL$ propagators respectively. Poles for these chiral propagators are clearly not seen.  

\section{Conclusion}
Using the gauge fixing approach  and
tuning only a finite number of counterterms (in this case, just the
$\kappa$-term), we end up in the reduced model with free domain wall
fermion theory consisting only of LH and RH chiral modes respectively at the
domain and the anti-domain wall. With the U(1) transverse gauge {\em dof} back
on the waveguide, only the RH fermions on the anti-domain wall will be gauged
(according to our construction). The gauge degrees of freedom are completely decoupled.

We reached our conclusions for the case of $y=1$ by performing a perturbation theory for the fermion propagators around $\tilde{\kappa}=\infty$ and comparing them to the numerical results at $\tilde{\kappa}=0.2$. The comparison is near perfect due to the robust properties of domain wall fermions (this does not happen nearly as nicely for the Wilson fermions) and shows that as long as $\tilde{\kappa}$ is big enough to avoid the FM-PM transition, it is as good as infinity, {\em i.e.}, the whole FM-FMD transition line is in the same universality class as the perturbation point $\tilde{\kappa}=\infty$.  

Investigations at $y\ll 1$ lead to the same qualitative conclusions and indicate that the  model for any nonzero Yukawa coupling belongs to one universality class. 

The transition to a nonabelian gauge group in
this gauge-fixing approach is nontrivial and should be pursued. A more detailed
account of our studies can be found in \cite{prd,plb}. 

The authors thank M.F.L. Golterman, Y. Shamir, P.B. Pal and K. Mukherjee for useful discussions. One of the authors (SB) thanks the Theory Group and Computer Section, SINP for providing facilities. 

\begin{figure}[t]
\begin{center}
\vspace{-1.3cm}
\hspace*{-0.3cm}\includegraphics[width=8.0cm,height=10.0cm]{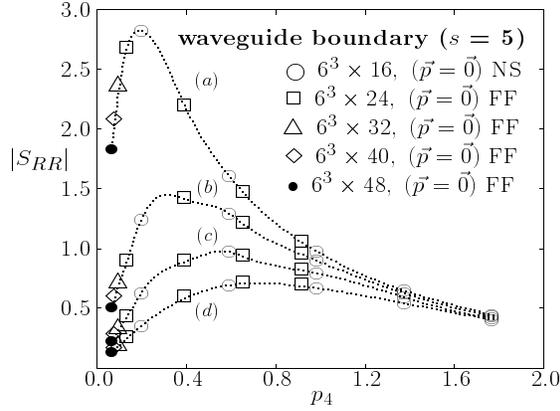}
\vspace{-4.0cm}
\caption{RR propagator at waveguide boundary $s=5$ with (a) $y=0.25$,
(b) $y=0.50$, (c) $y=0.75$, (d) $y=1.0$ ($L_s=22$; a.p.b.c. in $L_4$).}
\label{lr5y}
\end{center}
\vspace{-0.9cm}
\end{figure}

\begin{figure}[t]
\begin{center}
\vspace{-1.6cm}
\hspace*{-0.2cm}\includegraphics[width=8.0cm,height=10.0cm]{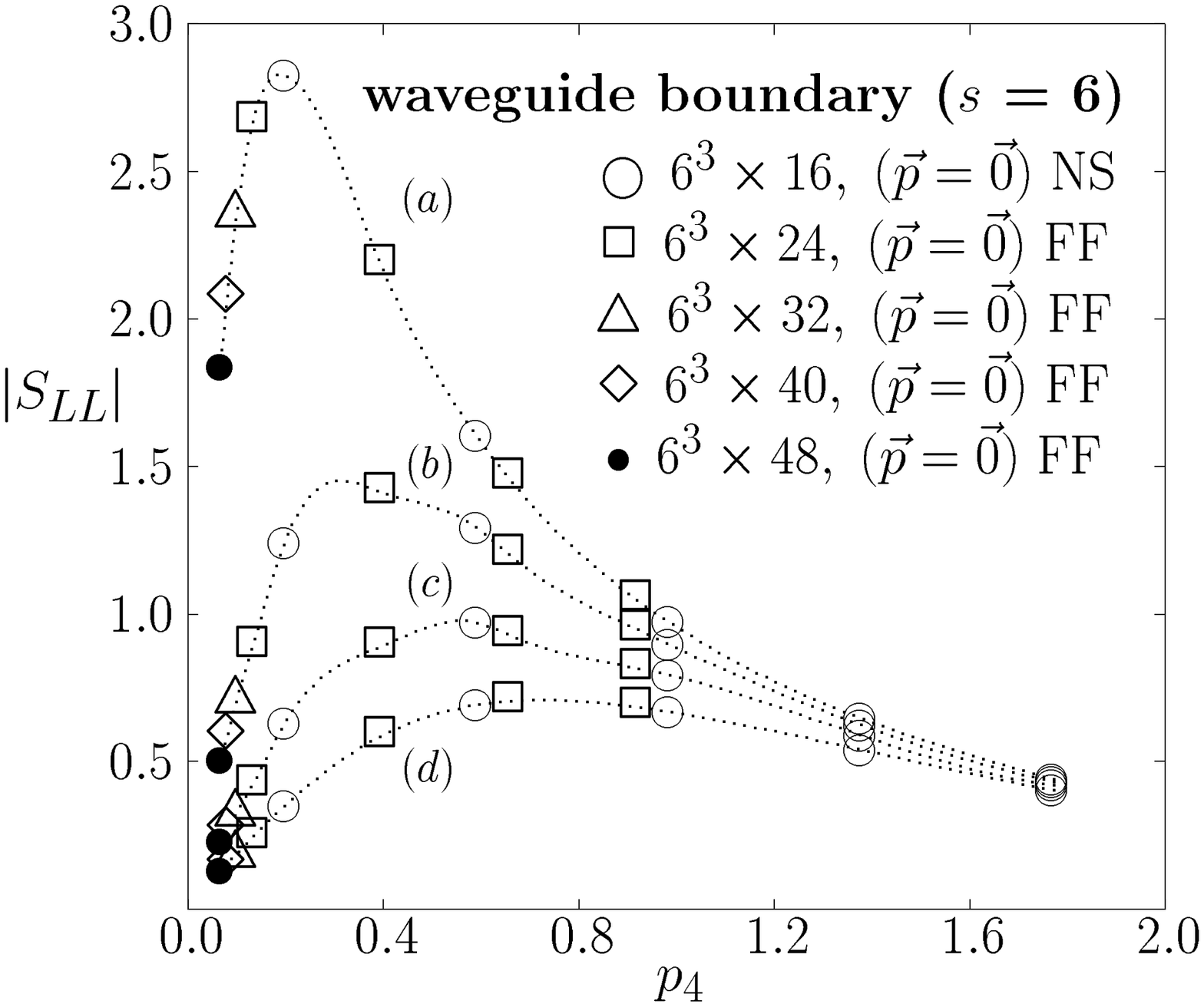}
\vspace{-4.0cm}
\caption{LL propagator at waveguide boundary $s=6$ with (a) $y=0.25$,
(b) $y=0.50$, (c) $y=0.75$, (d) $y=1.0$ ($L_s=22$; a.p.b.c. in $L_4$).}
\label{lr6y}
\end{center}
\vspace{-0.9cm}
\end{figure}

\end{document}